\documentclass[a4paper,11pt]{article}
\pdfoutput=1 

\usepackage{jheppub} 

\usepackage[utf8]{inputenc} 
\usepackage[T1]{fontenc} 

\usepackage{color}
\usepackage{amsmath}

\font\mybb=msbm10 at 12pt

\font\mybbsub=msbm10 at 8pt
\font\mybbsmall=msbm10 at 10pt

\def\bb#1{\hbox{\mybb#1}}

\def\bbsub#1{\hbox{\mybbsub#1}}
\def\bbsmall#1{\hbox{\mybbsmall#1}}

\def\FF {\bb{F}}
\def\FFsub{\bbsub{F}}
\def\FFsmall {\bbsmall{F}}

\def\ZZsmall {\bbsmall{Z}}

\def\PP {\bb{P}}
\def\PPsub{\bbsub{P}}
\def\PPsmall {\bbsmall{P}}

\newcommand\beqa{\begin{eqnarray}}
\newcommand\eeqa{\end{eqnarray}}
\newcommand\n{\nonumber\\}

\def\beq#1\eeq{\begin{equation}#1\end{equation}}
\def\bes#1\ees{\begin{equation}\begin{split}#1
               \end{split}\end{equation}}
\def\bea#1\eea{\begin{align}#1\end{align}}

\begin{document}

{~}

\title{Magic square and half-hypermultiplets \\in F-theory }

\author[a]{Rinto Kuramochi,}
\author[a,b]{Shun'ya Mizoguchi,}
\author[c]{Taro Tani}

\affiliation[a]{Graduate University for Advanced Studies (Sokendai)\\
Tsukuba, Ibaraki, 305-0801, Japan}
\affiliation[b]{Theory Center, Institute of Particle and Nuclear Studies, KEK\\
Tsukuba, Ibaraki, 305-0801, Japan}
\affiliation[c]{National Institute of Technology, Kurume College, \\Kurume, Fukuoka, 830-8555, Japan}
\emailAdd{rinto@post.kek.jp}
\emailAdd{mizoguch@post.kek.jp}
\emailAdd{tani@kurume-nct.ac.jp}

\abstract{In six-dimensional F-theory/heterotic string theory, 
half-hypermultiplets arise only when 
they correspond to particular 
quaternionic K\"ahler symmetric spaces, 
which are mostly associated with 
the Freudenthal-Tits magic square. 
Motivated by the intriguing singularity structure 
previously found in such F-theory models with a gauge group 
$SU(6)$, $SO(12)$ or $E_7$, 
we investigate, as the final magical example, 
an F-theory on an elliptic fibration over a Hirzebruch surface 
of the non-split $I_6$ type, 
in which the unbroken gauge symmetry is 
supposed to be $Sp(3)$. 
We find significant qualitative differences 
between the previous F-theory models associated with 
the magic square and   
the present case. 
We argue that the relevant half-hypermultiplets arise 
at the  $E_6$ points,
where half-hypermultiplets ${\bf 20}$
of $SU(6)$ would have appeared in the split model. 
We also consider the problem on the non-local matter 
generation near the $D_6$ point. After stating what the problem is, 
we explain why this is so by using the recent result that
a split/non-split transition can be regarded as a 
conifold transition.
}

\preprint{KEK-TH-2035}
\date{August 1, 2020}
\maketitle

\section{Introduction}
F-theory \cite{Vafa,MV1,MV2} is a framework of nonperturbative compactifications 
of type IIB string theory containing general $(p,q)$-7-branes. 
The nonperturbativeness of F-theory arises due to the nonlocality 
among the 7-branes and the strings, where 
the $SL(2,\ZZsmall)$ identification before and 
after a move of a string among 7-branes gives 
rise to open-string-like light pronged objects, string junctions. 
In the dual M-theory picture, they correspond to wrapped M2-branes 
around vanishing cycles. These objects account 
for the emergence of the exceptional gauge 
symmetry and matter in the spinor representation in a type II 
setup, which is one of the virtues of F-theory 
in the application to the phenomenological model building. 

In F-theory, 
matter typically arises at 
the intersections of 7-branes, where the singularity of 
the gauge brane with gauge group $H$ 
is ``enhanced'' to that labeled by some another  
higher-rank group $G$ \cite{MV1,MV2,BIKMSV,KatzVafa,Tani}.~\footnote{
The matter localization at the intersection 
of the spectral cover $C$ and the zero section $\sigma_{B_2}$
(in the 4D case) was originally shown in \cite{Curio,DiaconescuIonesei} 
by using the Leray spectral sequence. It is precisely where the 
singularity gets enhanced on $B_2$, though of course 
the spectral cover $C$ cannot be regarded as the matter 7-brane 
itself as it intersects with the elliptic fiber. This coincidence 
was explained in \cite{MTanomaly,MTLooijenga} 
in terms of the Mordell-Weil lattice 
of a rational elliptic surface \cite{OguisoShioda}. 
}
In generic cases, $G$ is one rank higher than $H$, and in six dimensions 
the matter arising at the intersection is 
in most cases a hypermultiplet transforming as 
$G/(H\times U(1))$, which determines a homogeneous K\"ahler manifold. 
However, in some cases, matter emerging at the intersection 
is not a full hypermultiplet but a {\em half-}hypermultiplet. 
For example \cite{BIKMSV},  
when $(G,H)$ are $(E_6,SU(6))$, $(E_7,SO(12))$ or $(E_8,E_7)$,
half-hypermultiplets in ${\bf 20}$, ${\bf 32}$ or ${\bf 56}$ 
of the respective $H$ appear.  
They are all pseudo-real representations 
and correspond, not to homogeneous K\"ahler manifolds,  
but to quaternionic K\"ahler symmetric spaces known as Wolf spaces
\cite{Wolf,Alekseevskii} (see \cite{Dasguptaetal} for a review):
\beqa
\frac{E_6}{SU(6)\times SU(2)},~~
\frac{E_7}{SO(12)\times SU(2)},~~
\frac{E_8}{E_7\times SU(2)}.
\label{threesymmetricspaces}
\eeqa

In \cite{MT}, an explicit resolution of the codimension-two singularity 
was carried out for the first example $(G,H)=(E_6, SU(6))$. 
It was found that the codimension-two singularity 
was already resolved by blowing up the nearby codimension-one 
$A_5=SU(6)$ singularities without any additional blow-up at  
that point, although the 
Kodaira fiber type right above the intersection point was $IV^*$, which 
would mean an $E_6$ singularity. 
The number of exceptional curves above the codimension-two 
point is the same as that of the codimension-one loci supporting 
a fiber of the type $I_6$. It was also found that 
the intersection diagram at the codimension-two point was different 
from that of the nearby codimension-one loci, explaining the 
generation of the half-hypermultiplet at that point. This type of 
resolution was called an {\em ``incomplete resolution''} \cite{MT}.
In \cite{KMT}, a similar analysis was performed for 
$(G,H)=(E_7, SO(12))$ and $(E_8, E_7)$ to find similar features. 

We should note that all these enhancements are relevant in 
the applications to F-theory GUT model buildings. For instance, the 
enhancement $SU(6)\rightarrow E_6$ is the one at the 
(codimension-three) Yukawa K\"ahler
point on the $\bar{\bf 5}$ matter curve in the four-dimensional 
$SU(5)$ F-GUT model. Similarly, the enhancements 
$SO(12)\rightarrow E_7$ and $E_7\rightarrow E_8$ are 
the ones at the Yukawa points on the $\overline{\bf 10}$ and ${\bf 27}$ 
curves in the $SO(10)$ and $E_6$ F-GUT models, respectively. 
Also, the multiple (=higher-rank) enhancement $SU(5)\rightarrow E_7$
(or $E_8$) (which includes these special enhancements as intermediate steps) 
is relevant to the F-theory family 
unification scenario \cite{FFamilyUnification} aiming to
implement the supersymmetric $E_7$ coset sigma model \cite{KugoYanagida} 
in F-theory.

Incidentally, the three symmetric spaces (\ref{threesymmetricspaces}) 
are precisely the ones obtained by taking a quotient of 
the groups of the entries of the Freudenthal-Tits 
magic square 
(Table \ref{magicsquare}).
The relation between quaternionic K\"ahler manifolds 
and the magic square was noticed some time 
ago in \cite{Dasguptaetal}. 
Indeed, the $G$'s and $H$'s comprising 
the symmetric spaces in (\ref{threesymmetricspaces}) 
are the groups of the Lie algebras listed in the bottom and 
the second bottom rows of the rightmost 
three columns in the table.
Motivated by this observation, in this paper 
we focus on the final remaining column of the 
magic square and study the corresponding six-dimensional F-theory 
compactification on an elliptic CY3 over a Hirzebruch surface \cite{MV1,MV2}. 
We can indeed find in \cite{BIKMSV} a model 
with the gauge group $C_3=Sp(3)$ yielding half-hypermultiplets 
in $F_4/(Sp(3)\times SU(2))={\bf 14}'$ as a part of the 
massless matter: the {\em non-split} $I_6$ model.  

One of our interests is what kind of singularity gives rise 
to the supermultiplet of chiral matter in this representation.
The equation defining the non-split $I_6$ 
model is obtained by modifying that of the split 
$I_6$ model \cite{BIKMSV}. 
The latter gives the $SU(6)$ unbroken gauge symmetry
with matter fields in ${\bf 20}$, ${\bf 15}$ and ${\bf 6}$ where the 
singularity is enhanced from $A_5$ to $E_6$, $D_6$ and $A_6$ 
respectively.  A ${\bf 20}$ is a half-hypermultiplet in the split case 
studied in \cite{MT}. 
We can obtain the equation for the non-split $I_6$ model 
by a certain change of the sections that characterize 
the equation of the split $I_6$ model. 
With this change, the local structures of the singularities at 
the $E_6$ and $A_6$ points remain intact, but only 
those at the $D_6$ points are affected, so 
we examine the singularity structure at  the $D_6$ points 
in the non-split $I_6$ model.

The non-spilt models are known to have some 
puzzles regarding the generation of matter fields \cite{BIKMSV,GHLST,AGW,EJK,EK,EJ}.
The equation defining the non-split $I_6$ 
model is obtained 
by replacing the square of a
 particular section $h_{n+2-r}^2$ (see text for 
 the definition) in the split 
$I_6$ equation with a non-square section $h_{2n+4-2r}$.
This global non-factorization implies monodromy among 
the exceptional fibers, which is 
interpreted as 
a feature that causes the gauge group to reduce 
from the simply-laced $SU(6)$ 
to the non-simply-laced $Sp(3)$ \cite{BIKMSV}. 
%
However, there is a puzzle here: 
At each double zero locus of $h_{n+2-r}$ there appears 
a hypermultiplet in ${\bf 15}$ of $SU(6)$ in the split model.
Therefore, the anomaly cancellation requires that the hypermultiplets 
in ${\bf 15}$ of $SU(6)$ at the double zeros 
should {\em split in pairs} according to the replacement of the section,   
but the ${\bf 14}$ (not ${\bf 14}'$ - see below) 
of $Sp(3)$, supposed
to arise from the ${\bf 15}$ of $SU(6)$, is a {\em real}
(not a {\em pseudo-real}) representation, which does 
not allow half-hypermultiplets.
This is the first puzzle.

There is another curious feature about this non-split model:
As in \cite{MT,KMT}, we consider 
a local equation which exhibits the singularity structure near a single 
zero locus of the section $h_{2n+4-2r}$. 
The resolution of the singularity turns out to be an 
``incomplete'' resolution, meaning that the codimension-two 
``$D_6$'' singularity is already resolved when the resolution of 
the codimension-one singularity is completed. However, 
the difference from the previous three magical examples is that 
the intersection matrix of the exceptional curves at 
the codimension-two \footnote{Note that this codimension 
is counted in the base space of the elliptic fibration, and 
not in the total space of the Calabi-Yau.} $D_6$
point remains identical to that at a 
nearby point on the codimension-one singularity.
Therefore, the configuration of the exceptional curves generated 
there does not indicate that any chiral matter field is localized there.

%
These puzzles require a new understanding of charged matter 
generation in the non-split model, other than 
wrapped branes around vanishing cycles \cite{KatzVafa} or 
string junctions ending on the intersections of 7-branes \cite{Tani}.
Very recently, 
it was shown \cite{KuMT} 
that the split/non-split transition in F-theory 
can be regarded as, except some exceptional cases, a conifold 
transition associated with the relevant conifold singularities.   
In this paper, we will use this fact to discuss how the necessary 
matter can emerge from the geometry of the non-split model.  
More precisely, since the non-split model corresponds to 
the ``deformed side'' of the conifold transition, there arise 
three-cycles instead of two cycles on the ``resolved side'', 
which is the split model. We will argue what branes can give 
chiral matter field with the three-cycles.

On the other hand, as for the 
question of where the ${\bf 14}'$s are generated, 
we argue that they just arise as the $Sp(3)$ decomposition 
of ${\bf 20}$s of $SU(6)$ at the $E_6$ points, and 
not at the $D_6$ points.

The organization of this paper is as follows: In section 2, 
we give a brief review of the Freudenthal-Tits magic square and 
point out its relation to half-hypermultiplets in F-theory. In section 3, 
we consider the global split and non-split 
$I_6$ models and examine their matter 
spectra.  
In section 4, we perform a concrete 
blowing-up process of the ``$D_6$'' singularity
of the non-split $I_6$ local equation. 
In section 5, we introduce the recent result of \cite{KuMT} 
and show how it is used to resolve the issue of non-local matter. 
The final section is devoted to conclusions.

\section{Magic square and half-hypermultiplets in F-theory}

\subsection{The Freudenthal-Tits magic square}
A Freudenthal-Tits magic square is a four-by-four table whose 
entries are Lie algebras. They are determined by specifying 
a pair of composition algebras $(\mathbb{A}, \mathbb{B})$.  
When these composition algebras are the ones over the real 
number field $\mathbb{R}$, they are either one of the four 
division algebras  $\mathbb{R}$, $\mathbb{C}$, $\mathbb{H}$ and 
$\mathbb{O}$, or they are one of the ``split'' algebras of $\mathbb{C}$, 
$\mathbb{H}$ and $\mathbb{O}$, which are non-compact analogues 
of the corresponding division algebras. In this case, each entry 
of the magic square 
is some real form of a complex Lie algebra.

If $(\mathbb{A}, \mathbb{B})$ are a pair of either of 
the four division algebras 
$\mathbb{R}$, $\mathbb{C}$, $\mathbb{H}$ and $\mathbb{O}$,
the magic square consists of compact Lie algebras with 
definite signatures (Table {\ref{magicsquare}}), while if $(\mathbb{A}, \mathbb{B})$ are chosen from the set of $\mathbb{R}$ and the 
three split algebras, the entries are all split real forms of the same 
complexifications as those of the compact Lie algebras in the 
corresponding cells. They typically arise (besides a few exceptions) 
as (Lie algebras of) duality groups or hidden symmetries of 
dimensionally reduced maximally symmetric supergravities, 
bosonic string or the NS-NS sector effective theory and pure gravities. 
Finally, if $\mathbb{A}$ is a division algebra and 
$\mathbb{B}$ is a split algebra, 
the magic square conprises a special set of real forms of 
exceptional Lie algebras arising as scalar manifolds of 
dimensional reductions of 
$D=5$ ``magical'' supergravities \cite{magical1,magical2,Kanmagical,Fukuchimagical}.

The $(\mathbb{A}, \mathbb{B})$ entry of the magic square 
always has the following structure:
\beqa
\mathfrak{der}\,\mathbb{A}\,\oplus\,
\mathfrak{der}\,\mathfrak{J}^{\mathbb{B}}\,\oplus\,
(\mathbb{A}_0\otimes\mathfrak{J}^{\mathbb{B}}_0),
\eeqa 
where $\mathfrak{der}\,\mathbb{A}$ and 
$\mathfrak{der}\,\mathfrak{J}^{\mathbb{B}}$ are 
the Lie algebras of the automorphism groups of 
$\mathbb{A}$ and $\mathfrak{J}^{\mathbb{B}}$, respectively, 
and $\mathbb{A}_0$ and $\mathfrak{J}^{\mathbb{B}}_0$ 
denote their traceless parts. 

For example, 
for the compact case
$\mathbb{A}, \mathbb{B}=\mathbb{R}, \mathbb{C}, \mathbb{H}, 
\mathbb{O}$ (Table {\ref{magicsquare}}), \footnote{In this paper, 
we use the notations $\mathfrak{sp}(n)$ and $Sp(n)$ to 
denote the Lie algebra and the Lie group of the $C_n$ type 
Dynkin diagram.}
\beqa
\mathfrak{der}\,\mathbb{A}&=&
0,0,\mathfrak{su}(2),\mathfrak{g}_2,\\
\mathfrak{der}\,\mathfrak{J}^{\mathbb{B}}&=&
\mathfrak{so}(3),\mathfrak{su}(3),\mathfrak{sp}(3),\mathfrak{f}_4,\\
\mathbb{A}_0&=&0,0,{\bf 3},{\bf 7}~~~\mbox{of $\mathfrak{der}\,\mathbb{A}$} ,\\
\mathfrak{J}^{\mathbb{B}}_0&=&{\bf 5},{\bf 8},{\bf 14},{\bf 26}~~~\mbox{of $\mathfrak{der}\,\mathfrak{J}^{\mathbb{B}}$}.
\eeqa
Then, for instance, $\mathfrak{e}_7$ allows a decomposition
\beqa
E_7&\supset&SU(2)\times F_4\n
{\bf 133}&=&({\bf 3},{\bf 1})\oplus({\bf 1},{\bf 52})
\oplus({\bf 3},{\bf 26})
\eeqa 
for $\mathbb{A}=\mathbb{H}, \mathbb{B}=\mathbb{O}$, 
and also 
\beqa
E_7&\supset&G_2\times Sp(3)\n
{\bf 133}&=&({\bf 14},{\bf 1})\oplus({\bf 1},{\bf 21})
\oplus({\bf 7},{\bf 14})
\eeqa 
for $\mathbb{A}=\mathbb{O}, \mathbb{B}=\mathbb{H}$.
The other Lie algebras allow similar decompositions.

\begin{table}[tbp]
\centering
\begin{tabular}{|c||c|c|c|c|}
\hline 
$\mathbb{B}\backslash\mathbb{A}$&
$\mathbb{R}$&$\mathbb{C}$&$\mathbb{H}$&$\mathbb{O}$\\
\hline 
$\mathbb{R}$ &$\mathfrak{so}(3)$&$\mathfrak{su}(3)$
&$\mathfrak{sp}(3)$&$\mathfrak{f}_4$\\
$\mathbb{C}$ & $\mathfrak{su}(3)$ & $\mathfrak{su}(3)\oplus\mathfrak{su}(3)$&$\mathfrak{su}(6)$&$\mathfrak{e}_6$\\
$\mathbb{H}$ &$\mathfrak{sp}(3)$&$\mathfrak{su}(6)$
&$\mathfrak{so}(12)$&$\mathfrak{e}_7$\\
$\mathbb{O}$ &$\mathfrak{f}_4$&$\mathfrak{e}_6$
&$\mathfrak{e}_7$&$\mathfrak{e}_8$\\
\hline
\end{tabular}
\caption{\label{magicsquare} The Freudenthal-Tits magic square 
for  $\mathbb{A}, \mathbb{B}$ being either of the four 
division algebras $\mathbb{R}, \mathbb{C}, \mathbb{H}, 
\mathbb{O}$. They are all compact Lie algebras with definite 
signatures. If the division algebras are replaced by split composition 
algebras, the entries become different real forms with the same 
complexifications.}
\end{table}
\vskip 2ex
\noindent
{\it Remark.} In this paper the word ``split'' is used in three
different meanings:
\begin{itemize}
\item[1.]{This word is used for a ``split'' composition algebra, which 
is a noncompact version 
of $\mathbb{C}$, $\mathbb{H}$ or $\mathbb{O}$ 
with an indefinite bilinear form.}
\item[2.]{``Split'' is also used for a ``split'' real form of a
complex Lie algebra, which has, besides the Cartan subalgebra, 
an equal number of positive and negative generators with 
respect to the invariant bilinear form.}
\item[3.]{Finally, the word ``split'' appears in the classification 
of singularities or the fiber types of exceptional curves 
\cite{BIKMSV}. 
Singularities of the ``split'' type are 
the ones in which relevant exceptional curves factor globally 
so that they yield simply-laced gauge symmetries. }
\end{itemize}
The first two are closely related in that split real forms of the 
item 2 arise in the magic square when the composition algebras 
are taken to be split ones in the sense of item 1.  
The third one is, however, a different notion from the two.

\subsection{Half-hypermultiplets in F-theory}
In \cite{BIKMSV}, a detailed analysis was carries out 
on the matter spectra of six-dimensional F-theory compactifications 
on an elliptically fibered Calabi-Yau threefold over a Hirzebruch surface \cite{MV1,MV2} for various patterns of unbroken gauge groups.
In particular, it was revealed that there were (essentially) four cases 
of unbroken gauge groups %
\footnote{There is, in fact, one more example in 
\cite{BIKMSV} 
where half-hypermultiplets arise as massless matter: 
the ${\bf 32}$ of $SO(11)$. 
This is also a non-split model ($I_2^{*ns}$), and 
this ${\bf 32}$ is easily seen to arise 
at the $E_7$ point, where the corresponding split 
model  ($I_2^{*s}$) with the $SO(12)$ gauge symmetry 
also yields ${\bf 32}$.
} 
in which {\em half-hypermultiplets} (rather than normal hypermultiplets) 
appeared as massless matter. They are listed in Table \ref{halfhyper} and 
\ref{halfhyper2}. These spectra can be confirmed either by 
the heterotic index calculation \cite{GSW}~\footnote{
For $Sp(3)$, the dual heterotic gauge bundle is $SU(2)\times G_2$
since the maximal embedding is $E_8 \supset SU(2)\times G_2 \times Sp(3)$
(see {\it e.g.} \cite{Yamatsu} for the branching rules).
The spectrum in Table \ref{halfhyper2} is obtained by distributing 
the $12+n$ instantons as $(4+r,8+n-r)$ in $(SU(2),G_2)$.}
or by the generalized 
Green-Schwarz mechanism using the divisor data of the 
Hirzebruch surface \cite{Sadov,MizoguchiTanianomaly}.~\footnote{
For $Sp(3)$, the relevant indices of  a representation ${\bf R}$ for examining 
the generalized Green-Schwarz (GS) mechanism are 
given by $(\mbox{index}({\bf R}), x_{\bf R}, y_{\bf R})=(8,14,3)$, $(1,1,0)$, $(4,-2,3)$ and $(5,-7,6)$ for 
${\bf R}={\bf Adj}$, ${\bf 6}$, ${\bf 14}$ and ${\bf 14}'$, respectively, where 
$\mbox{tr}_{\bf R}F^2=\mbox{index}({\bf R}) \mbox{tr}_{\bf 6}F^2$
and $\mbox{tr}_{\bf R}F^4=x_{\bf R} \mbox{tr}_{\bf 6}F^4+y_{\bf R} (\mbox{tr}_{\bf 6}F^2)^2$.
By using these data and assuming that the charged matter spectrum only
contains ${\bf 6}$, ${\bf 14}$ and ${\bf 14}'$, one can solve the equations 
of generalized GS mechanism on $\FFsmall_n$ and 
obtain the unique solution given in Table \ref{halfhyper2}.}
They satisfy the anomaly free constraint for one of the $E_8$ factors
with instanton number $12+n$~\cite{BIKMSV}
\beq
 n_H-n_V = 30n+112.
\eeq

\begin{table}[tbp]
\centering
\begin{tabular}{|c|c|c|c|c|c|}
\hline 
$\begin{array}{cc}\mbox{gauge group}\\\mbox{$H$}
\end{array}$
&fiber type&
$\begin{array}{cc}\mbox{enhancement}\\\mbox{$G$}
\end{array}$
&matter rep.&multiplicity&
$\begin{array}{cc}\mbox{homogeneous}\\\mbox{space}
\end{array}$\\
\hline
$E_7$&$III^{*s}$&$E_8$&$\frac12${\bf 56}&$n+8$&
$\frac{E_8}{E_7\times SU(2)}$\\
&&&{\bf 1}&$2n+21$&$-$\\
\hline
$D_6$&$I_2^{*s}$&$E_7$&$\frac12${\bf 32}&$n+4$&
$\frac{E_7}{SO(12)\times SU(2)}$\\
&&$D_7$&{\bf 12}&$n+8$&
$\frac{SO(14)}{SO(12)\times U(1)}$\\
&&&{\bf 1}&$2n+18$&$-$\\
\hline
$A_5$&$I_6^s$&$E_6$&$\frac12${\bf 20}&$r$&
$\frac{E_6}{SU(6)\times SU(2)}$\\
&&$D_6$&{\bf 15}&$n+2-r$&
$\frac{SO(12)}{SU(6)\times U(1)}$\\
&&$A_6$&{\bf 6}&$2n+16+r$&
$\frac{SU(7)}{SU(6)\times U(1)}$\\
&&&{\bf 1}&$3n+21-r$&$-$\\
\hline
\end{tabular}
\caption{\label{halfhyper} Three cases in which half-hypermultiplets 
appear as massless matter in six-dimensional 
F-theory on an elliptic CY3 over $\mathbb{F}_n$ / 
heterotic string theory on K3 (quated  
from Table 3 of \cite{BIKMSV}). 
}
\end{table}
\begin{table}[htbp]
\centering
\begin{tabular}{|c|c|c|c|c|c|}
\hline 
\mbox{gauge group}&\mbox{representation}&\mbox{multiplicity}\\
\hline
$C_3$&$\frac12 ({\bf 14'}+{\bf 6})$&$r$\\
           &{\bf 14}&$n+1-r$\\
           &{\bf 6}&$2n+16+r$\\
           &{\bf 1}&$4n+23-2r$\\
\hline
\end{tabular}
\caption{\label{halfhyper2} The massless matter spectrum of 
six-dimensional heterotic string theory 
on K3 with an unbroken $Sp(3)$ gauge symmetry. 
This is anomaly free, and also contains half-hypermultiplets.
}
\end{table}

As we can see, the representations {\bf 56}, {\bf 32},
{\bf 20}, together with ${\bf 14}'$ and ${\bf 6}$, 
to which the half-hypermultiplets 
belong, are precisely the ones of quaternionic K\"ahler manifolds 
(or ``Wolf spaces''). All but the last ${\bf 6}$ are 
obtained by taking the Lie groups of the extreme bottom 
and the third rows of the magic square as the groups of 
the numerator and denominator of the homogeneous space.
The denominator groups also always come with an $SU(2)$ 
factor in contrast to the case of ordinary hypermultiplets, 
where the denominator group comprises not an $SU(2)$ 
but a $U(1)$ factor. In the latter case, the symmetric space is 
a homogeneous K\"ahler manifold \cite{FFamilyUnification}. 
In the M-theory Coulomb branch analysis of 
codimension-two or higher singularities \cite{BoxGraphs}, 
the Weyl-group invariant phases of this $SU(2)$ were shown 
to correspond to the resolutions yielding half-hypermultiplets. 


Let us summarize what is known so far,
for the 
three simply-laced split examples of Table \ref{halfhyper}, 
about the resolutions 
of the codimension-two singularities that yield half-hypermultiplets.
The resolutions of the third example were studied in \cite{MT}, 
and the those of the first and second ones were worked out 
in \cite{KMT}. 
The main relevant features are\footnote{The local coordinate $s$ 
parametrizing the base $\PPsub^1$ of $\FFsub_n$ will be 
denoted by $w$ in section 4 when we blow up 
the singularities.} :
\begin{itemize}
\item[(i)] As in \cite{MV1,MV2}, 
let $z$ ($z'$) be the affine coordinate of the $\mathbb{P}^1$
fiber ($\mathbb{P}^1$ base) of the Hirzebruch surface 
$\mathbb{F}_n$, respectively. Suppose that we have a codimension-one 
singularity along the line $z=0$ with the fiber type specified  
in the second column of Table \ref{halfhyper}. 
Non-singlet matter arises where the singularity is ``enhanced'' 
from $H$ to $G$, in the sense that the Kodaira fibers  
read off at {\em right above that point} have 
intersections specified by the Dynkin diagram of $G$. 
However, where the half-hypermultiplets appear, the 
codimension-two singularity is already resolved 
by blowing up the nearby codimension-one singularities. 
No additional blow-up at the codimension-two point 
is required, even though the singularity is ``enhanced'' there
in the sense explained above. Such type of resolution is called 
an {\em incomplete resolution} \cite{MT}.

\item[(ii)] In an incomplete resolution, the relevant section 
that vanishes at codimension two goes like $O(s)$, where $s$ 
is a local coordinate holomorphic in $z'$,  
and $s=0$ is the codimension-two singularity. In this case, 
although the number of blow-ups required to resolve it 
is the same as that to resolve the nearby generic codimension-one 
singularities, the intersection matrix of the exceptional curves 
at $s=0$ is not the same as the generic one 
determined by the Cartan matrix of $H$ (nor that of $G$), but 
turns out to be a curious non-Dynkin diagram with some 
nodes having self-intersections $-\frac32$.

\item[(iii)] In the first three examples of Table \ref{halfhyper} 
studied in \cite{MT} and \cite{KMT}, 
$\frac32$ is the length square of the weight vector of 
the representations to which the half-hypermultiplets belong. 
It was confirmed that although the intersection matrix 
was not the (minus of the) Cartan matrix of $G$, 
the exceptional curves at $s=0$ formed an extremal ray that 
could span all the weights of the relevant 
pseudo-real representation of the half-hypermultiplets.  

\item[(iv)] In the first two examples, there arise several codimension-one 
singularities during the intermediate stages of the blow-up 
process, and there are several options in 
which singularity we blow up first, and which we do 
afterwards. 
Depending on the ordering of the blow-ups, one obtains 
different intersection diagrams of the exceptional curves 
at the codimension-two point $s=0$ \cite{KMT}. 
More specifically,  the intersection diagram 
on every other row found in \cite{BoxGraphs} 
can be obtained in this way, 
but not all of them. 

\item[(v)] Instead, when the relevant section vanishes like $O(s^2)$ 
at the codimension-two point, the singularity becomes stronger 
than the case above 
so that there arises an additional conifold singularity. A small 
resolution generates an extra exceptional fiber at that point so 
that it completes the proper Dynkin diagram of group $G$. 
This type of resolution is called a {\em complete resolution} 
\cite{MT}. 
\end{itemize}

\section{Six-dimensional $Sp(3)$ global model}
\subsection{The non-split $I_6$ equation on $\FF_n$}
In this section we consider a six-dimensional F-theory compactification 
on an elliptic fibration over a Hirzebruch surface $\mathbb{F}_n$ 
in which the unbroken gauge symmetry reduces to $Sp(3)$. 
We work in the $dP_9$ fibration so that we focus on one of the 
two $E_8$'s  of the heterotic dual.

As was shown in \cite{BIKMSV}, the equation of this curve is 
the one that supports a $I_6$ Kodaira fiber of the non-split 
type at $z=0$.
A $I_6$ non-split curve may be obtained by replacing 
the relevant factorized section of a split $I_6$ curve 
with a non-factorized one. More specifically,
consider Tate's form of the equation describing the elliptic fibration:
\beqa
-(y^2+a_1 x y + a_3 y)+x^3+ a_2 x^2 + a_4 x + a_6=0.
\label{Tate's}
\eeqa
As in \cite{MV1,MV2}, we use $z$ and $z'$ as the affine 
coordinates of the base and fiber $\PPsmall^1$'s of the 
Hirzebruch surface. 
The equation for the theory with the unbroken group $H=SU(6)$ 
can be obtained by specializing the sections as
\beqa
a_1&=& 2 \sqrt{3} t_r h_{n-r+2}, \n
a_2&=&
   -3 z t_r H_{n-r+4},\n
   a_3&=& 2 \sqrt{3} z^2 u_{r+4}
   h_{n-r+2},\n
   a_4&=& z^3 \left(t_r f_{n-r+8}-3
   u_{r+4} H_{n-r+4}\right)+f_8 z^4,\n
   a_6&=& z^5
   u_{r+4} f_{n-r+8}+g_{12} z^6,
\label{SU(6)specializtion}
\eeqa
where $t_r$, $h_{n-r+2}$, $H_{n-r+4}$, $u_{r+4}$ and 
$f_{n-r+8}$ (together with $f_8$ and $g_{12}$) 
are the sections of appropriate 
line bundles over the base $\PPsmall^1$ 
specified by their subscripts, which in this case 
denote nothing but the degrees of the polynomials in $z'$.
It can be  
verified that the equation (\ref{Tate's})
with (\ref{SU(6)specializtion}) correctly reproduces the 
anomaly-free heterotic massless spectrum for an unbroken 
$SU(6)$ gauge group with $SU(3)\times SU(2)$ 
instanton numbers $(r, 12+n-r)$ (see e.g.\cite{MTnonCartan}).
\vskip 2ex
\noindent
{\it Remark.}  While (\ref{Tate's}) and (\ref{SU(6)specializtion})
successfully yields a consistent $SU(6)$ model, the vanising orders 
of $(a_1, a_2, a_3, a_4, a_6)$ in $z$ are $(0,1,2,3,5)$, which are 
the same as those for the split $I_5$ fiber type $I_5^s$ 
and differ from 
the ``standard'' Tate's orders $(0,1,3,3,6)$ 
for the split $I_6$ fiber type $I_6^s$ 
classified in \cite{BIKMSV}. Indeed, it can be easily seen 
that the sections $(a_1, a_2, a_3, a_4, a_6)$ with orders 
$(0,1,3,3,6)$ only result in the Weierstrass model 
(\ref{WeierstrassSU(6)})(\ref{fSU(6)})(\ref{gSU(6)})
with constant $t_r$, that is, no instantons are 
distributed to the $SU(3)$ factor, and all the $12+n$ 
instantons are in the $SU(2)$ factor. In fact, one can redefine $y$ 
and $x$ so that the vanishing orders of 
$(a_1, a_2, a_3, a_4, a_6)$ may become $(0,1,3,3,6)$ 
only when $t_r\neq 0$, but  cannot when $t_r=0$ 
since the redefinitions of $y$ and $x$ contain shifts 
proportional to $\frac1{t_r}$, which diverge at $t_r=0$.


By redefining $y$ and $x$, we obtain the Weierstrass equation
\beqa
0&=&-y^2+x^3+f_{SU(6)}(z,z')x + g_{SU(6)}(z,z'),
\label{WeierstrassSU(6)}\\
f_{SU(6)}(z,z')&\equiv&
-3 t_r^4 h_{n-r+2}^4
+6z  t_r^3 h_{n-r+2}^2 
   H_{n-r+4} \n&&
   +z^2 \left(6 t_r u_{r+4} h_{n-r+2}^2-3 t_r^2
   H_{n-r+4}^2\right)\n
   &&+z^3 \left(t_r f_{n-r+8}-3 u_{r+4}
   H_{n-r+4}\right)
   +f_8 z^4,
   \label{fSU(6)}
   \\
g_{SU(6)}(z,z')&\equiv&
2 t_r^6 h_{n-r+2}^6-6 z \left(t_r^5 h_{n-r+2}^4
   H_{n-r+4}\right)\n
   &&-6 z^2 \left(t_r^3 u_{r+4}
   h_{n-r+2}^4-t_r^4 h_{n-r+2}^2 H_{n-r+4}^2\right)\n
   &&+z^3
   \left(-t_r^3 f_{n-r+8} h_{n-r+2}^2+9 t_r^2 u_{r+4}
   h_{n-r+2}^2 H_{n-r+4}-2 t_r^3 H_{n-r+4}^3\right)\n
   &&+z^4
   \left(-f_8 t_r^2 h_{n-r+2}^2+t_r^2 f_{n-r+8}
   H_{n-r+4}+3 u_{r+4}^2 h_{n-r+2}^2-3 t_r u_{r+4}
   H_{n-r+4}^2\right)\n
   &&+z^5 \left(f_8 t_r H_{n-r+4}+u_{r+4}
   f_{n-r+8}\right)+g_{12} z^6
   \label{gSU(6)}
\eeqa
with a discriminant
\beqa
4f_{SU(6)}^3+27g_{SU(6)}^2&=&
z^6 t_r^3 h_{n-r+2}^4 
P_{2n+r+16}
~
+ z^7 t_r^2 h_{n-r+2}^2 
Q_{3n+20}
~
+z^8 
R_{4n+24}
\n
   &&+O(z^9),
   \label{DeltaSU(6)}
\eeqa
where $P_{2n+r+16}$, $Q_{3n+20}$ and $R_{4n+24}$ 
are some non-factorizable polynomials 
in $z'$ of degrees specified by the subscripts.
In generic cases, any two of $t_r$,  $h_{n-r+2}$ and  
$P_{2n+r+16}$ do not share a common zero locus, which 
we assume in this paper.
From (\ref{fSU(6)}), (\ref{gSU(6)})
and  (\ref{DeltaSU(6)}) we can see that the Kodaira fiber types 
above the zero loci of $t_r$,  $h_{n-r+2}$ and  
$P_{2n+r+16}$ are respectively $IV^*$, $I_2^*$ and $I_7$, 
yielding the singularity enhancements from $H=SU(6)$ 
to $G=E_6$, $D_6$ and $A_6$ 
as presented in the third column of 
Table \ref{halfhyper}. 
We can also see that the $h_{n-r+2}$-dependence of 
$f_{SU(6)}$ (\ref{fSU(6)}) or $g_{SU(6)}$ (\ref{gSU(6)}) 
is only through $h_{n-r+2}^2$, which allows us to replace 
every $h_{n-r+2}^2$ in  $f_{SU(6)}$ and $g_{SU(6)}$ 
with a generic polynomial $h_{2n-2r+4}$. The resulting 
equation is the one for $I_6^{ns}$ \cite{BIKMSV}.

\subsection{The massless spectrum}
As we will see explicitly in the next section, the replacement 
of the section $h_{n-r+2}^2\rightarrow h_{2n-2r+4}$ in the 
split $I_6$ equation results in the global non-factorization of 
the exceptional curves, which reduces the 
gauge group from $SU(6)$ to $Sp(3)$.  
Let us examine what matter multiplets are 
expected to arise in this model.

In the transition $I_6^s\leftrightarrow I_6^{ns}$, 
nothing changes
in the local singularity structure near the zero loci of $t_r$ and $P_{2n+r+16}$,
where $\frac12{\bf 20}$ and {\bf 6} of $SU(6)$ appear as massless matter
in the split theory; 
the string junctions or the vanishing cycles there do not ``know'' 
whether the total equation is of the split type or of the non-split type. 
The only change they feel is that of the gauge group, so
they simply decompose into irreducible representations 
of $Sp(3)$, which is the gauge group of the non-split theory.
Thus, at a zero locus of $t_r$, a half-hypermultiplet in {\bf 20} of $SU(6)$, 
of which the quaternionic K\"ahler manifold 
$E_6/(SU(6)\times SU(2))$ is comprised, 
is decomposed into half-hypermultiplets in ${\bf 14}'$ and ${\bf 6}$ of 
$Sp(3)$, while at a zero of $P_{2n+r+16}$, a hypermultiplet 
in {\bf 6} of $SU(6)$ entirely becomes one in {\bf 6} of $Sp(3)$.
Note that {\bf 6} is also a pseudo-real representation of $Sp(3)$, 
and the latter can be regarded as $2n+r+16$ pairs of 
half-hypermultiplets. The ${\bf 14}'$ constitutes 
the quaternionic K\"ahler manifold 
$F_4/(Sp(3)\times SU(2))$, while the ${\bf 6}$ does 
$Sp(4)/(Sp(3)\times SU(2))$. This will answer to the original question 
of where the matter fields corresponding to the final magical 
coset arise; they arise at the $E_6$ points of the non-split 
$I_6$ model as an irreducible multiplet in the 
$Sp(3)$ decomposition of ${\bf 20}$ of $SU(6)$.

\subsection{A puzzle on matter fields near the $D_6$ points}
On the other hand, there is a puzzle as we mentioned in Introduction: 
With the replacement $h_{n-r+2}^2\rightarrow h_{2n-2r+4}$,  
the $n-r+2$ double roots of the equation $h_{n-r+2}^2=0$ split into 
$n-r+2$ pairs of single roots of $h_{2n-2r+4}=0$. Thus 
the number of loci where hypermultiplets in {\bf 15} of $SU(6)$ occur 
are doubled. A {\bf 15} of $SU(6)$ decomposes into ${\bf 14}\oplus{\bf 1}$
(and not ${\bf 14}'\oplus{\bf 1}$) of $Sp(3)$.
Since the adjoint of $SU(6)$ decomposes as
${\bf 35}={\bf 21}\oplus{\bf 14}$, where ${\bf 21}$ is the adjoint of $Sp(3)$, 
one ${\bf 14}$ of $n-r+2$ hypermultiplets can be thought of as 
eaten by the $SU(6)$ vector multiplet. 
Thus 
the anomaly-free massless matter spectrum  
shown in Table.\ref{halfhyper2} can be reproduced if the 
$n-r+2-1$ hypermultiplets in ${\bf 14}$ 
are ``distributed'' at the $2n-2r+4$ zero loci of $h_{2n-2r+4}$.
This, however, seems impossible, since the ${\bf 14}$ of $Sp(3)$ 
is a real representation and does not allow half-hypermultiplets 
in this representation. 

Of course, the original $SU(6)$ spectrum is already anomaly free, so 
hypermultiplets in ${\bf 14}$ can not be present equally at all the $2n-2r+4$ 
zeros of $h_{2n-2r+4}=0$ as they are too many to be anomaly free.
If they were ${\bf 14}'$ instead of ${\bf 14}$, 
they could be split into pairs and 
equally be distributed (up to the eaten ones) at the $2n-2r+4$ 
zeros, but both the heterotic anomaly analysis and Sadov's generalized
anomaly cancellation mechanism tell us that they must be ${\bf 14}$,
and not ${\bf 14}'$.

This poses a question of how the $n-r+1$ matter in ${\bf 14}$
of $Sp(3)$ are generated and where they reside 
in the non-split $I_6$ model.
In the next section, in order to explore what happens near a zero locus 
of $h_{2n-2r+4}$, we perform an explicit blow-up of the singularity. 

\section{Resolutions of the singularities}
\subsection{The local equation}
In this section, we carry out the process of blow-up of the codimension-two 
singularity at a zero locus of $h_{2n-2r+4}=0$.
To this aim, we consider a local equation 
in which the enhancement  of ``$A_5$'' to ``$D_6$''  
is achieved at codimension two.
\footnote{Again, they are quoted because they only imply the Lie
algebras whose Dynkin diagrams specify the intersections 
of the Kodaira fibers right above those points with fixed $z'$.} 
To obtain such an equation, 
We first complete the square with respect 
to $y$ in (\ref{Tate's}) and substitute (\ref{SU(6)specializtion}) into it. 
Writing $y+\frac12(a_1 x + a_3) \equiv Y$, we 
have 
\beqa
&&
-Y^2+x^3
+x^2 \left(3 t_r^2 h_{n-r+2}^2-3 z
   t_r H_{n-r+4}\right)
   \n
&&
+x \left(z^3 t_r f_{n-r+8}+f_8 z^4+6 z^2 t_r u_{r+4} h_{n-r+2}^2-3 z^3 u_{r+4}
   H_{n-r+4}\right)
   \n
   &&+3 z^4 u_{r+4}^2 h_{n-r+2}^2
   +z^5 u_{r+4} f_{n-r+8}+g_{12} z^6
   ~~=~~0,
\label{SU(6)b-form}
\eeqa
in which $h_{n-r+2}$'s appear only in the form $h_{n-r+2}^2$.
Thus we can make a replacement $h_{n-r+2}^2\rightarrow h_{2n-2r+4}$
in (\ref{SU(6)b-form}). 
By setting\footnote{
In this section, the local coordinates of the base $\PPsub^1$ 
of $\FFsub_n$ (whose affine coordinate is $z'$) 
will be denoted by $w$ and not by $s$, 
in accordance with \cite{KuMT}.} 
\beqa
h_{n-r+2}^2\rightarrow h_{2n-2r+4}&=& w,\n
t_r=H_{n-r+4}=u_{r+4}&=& \frac{1}{\sqrt{3}},\n
f_{n-r+8}=f_8=g_{12}&=& 0,
\eeqa
we can obtain a desired equation, but it is more convenient to 
make a shift in the $x$ coordinate $x+z^2\equiv X$. 
In terms of $X$, the final equation is
\beqa
-Y^2+X^3+X^2 \left(w- z (3 z+1)\right)+X (3 z+1) z^3- z^6=0,
\label{equationweblowup}
\eeqa
which we blow up in the follwing section.

If we write (\ref{equationweblowup}) as 
\beqa
-Y^2+X^3+\frac{b_2}4 X^2+\frac{b_4}2 X+\frac{b_6}4=0,
\eeqa
the vanishing orders of the sections $b_2$, $b_4$, $b_6$ in $z$ 
are $0$, $3$, $6$, respectively, which satisfy the criteria for 
the $I_6$ type Kodaira fiber in Tate's algorithm. This is due to 
the shift $x+z^2\equiv X$, as without it one would have instead 
the vanishing orders $0$, $2$, $4$. 
Note that such a shift of the variable $x$ 
to eliminate the order-$2$ term in $z$ from $b_4$  
is not possible globally, since
near a zero locus of $t_r$, where a $\frac12{\bf 20}$ 
of $SU(6)$ (or $\frac12({\bf 14'}\oplus {\bf 6})$ of $Sp(3)$) 
appears, the necessary shift becomes divergent. 
This is why an equation with 
$\mbox{ord}(b_2,b_4,b_6)=(0,2,4)$ was used in 
\cite{MT,KMT}.

\subsection{Blowing up the singularity}
Let us now consider the resolution of the singularity of 
the local equation (\ref{equationweblowup})
\beqa
\Phi(x,y,z,w)&\equiv&
-y^2+x^3+x^2 \left(w- z (3 z+1)\right)+x (3 z+1) z^3- z^6~=~0,
\label{equationweblowupXYwithxy}
\eeqa
where we have replaced $X,Y$ with $x,y$.
The equation (\ref{equationweblowupXYwithxy}) 
has a codimension-one singularity along $(x,y,z)=(0,0,0)$ 
for arbitrary $w$. 
\\
\\
\noindent
{\bf 1st blow up} \\
As was done in the previous works, we replace the complex line 
$(x,y,z)=(0,0,0)$ with $\mathbb{P}^2\times {\mathbb C}$ in ${\mathbb C}^4$
and examine the singularities of the local equations in three different 
charts corresponding to the affine patches of the $\mathbb{P}^2$ for 
some fixed $w$.
We also give the explicit forms of the exceptional curves $\cal C$'s 
at $w \neq 0$ and $\delta$'s at $w=0$.
($\delta$ is defined by the $w\rightarrow 0$ limit of $\cal C$ in the 
chart where $\cal C$ arises.)
\\
\\
\noindent
\underline{Chart $1_x$}
\beqa
\Phi(x, x y_1, x z_1,w)&=&x^2\Phi_x(x,y_1,z_1,w),\n
\Phi_x(x,y_1,z_1,w)&=&
w-x^4 z_1^6+3 x^3 z_1^4+x^2 (z_1-3) z_1^2-x z_1+x-y_1^2.\n
  %
  \mbox{$\cal C$$^{\pm}_{p_1}$ in $1_x$}&:&x=0,~~ y_1=\pm \sqrt{w}.\n
  \mbox{$\delta$$_{p_1}$ in $1_x$}&:&x=0,~~ y_1=0.\n
  \mbox{Singularities}&:&\mbox{None}.
\label{1x}
\eeqa
\\
\noindent
\underline{Chart $1_y$}
\beqa
\Phi(x_1 y, y, y z_1,w)&=&y^2\Phi_y(x_1,y,z_1,w),\n
\Phi_y(x_1,y,z_1,w)&=&
w x_1^2+x_1^3 y-x_1^2 y z_1 (3 y z_1+1)+x_1 y^2 z_1^3 (3 y z_1+1)-y^4 z_1^6-1
.\n
  %
    \mbox{$\cal C$$^{\pm}_{p_1}$ in $1_y$}&:& y=0,~~x_1=\pm1/\sqrt{w}.\n
    \mbox{$\delta$$_{p_1}$ in $1_y$}&:& \mbox{Invisible}. \n
   \mbox{Singularities}&:&\mbox{None}.
\label{}
\eeqa
\\
\noindent
\underline{Chart $1_z$}
\beqa
\Phi(x_1 z, y_1 z, z,w)&=&z^2\Phi_z(x_1,y_1,z,w),\n
\Phi_z(x_1,y_1,z,w)&=&
w x_1^2+z \left(x_1^3-x_1^2 (3 z+1)+x_1 z (3 z+1)-z^3\right)-y_1^2.
\n
  %
   \mbox{$\cal C$$^{\pm}_{p_1}$ in $1_z$}&:&z=0,~~y_1=\pm \sqrt{w} x_1.\n
   \mbox{$\delta$$_{p_1}$ in $1_z$}&:& z=0,~~y_1=0. \n
  \mbox{Singularities}&:&(x_1,y_1,z)=(0,0,0).
\label{}
\eeqa

Here, the chart $1_x$ is 
the affine patch of $\mathbb{P}^2 \ni(x:y:z)$ 
for $x\neq 0$ in which $(x:y:z)=(1:y_1:z_1)$. 
The other charts are also similar.\footnote{Note that we have 
used the same ``$z_1$'' in $1_x$ and $1_y$ for different coordinate variables, and similarly for $x_1$ and $y_1$. There will be no confusion 
as we do not compare equations in different charts. }
\\
\\
\noindent
{\bf 2nd blow up} \\
As we can see, the only singularity after the first blow up is 
$(x_1,y_1,z)=(0,0,0)$ on the chart $1_z$, 
which is not visible from the other charts. This is codimension one, 
and we blow up this singularity by similarly inserting a one-parameter ($=w$) 
family of $\PP^2$ along  $(x_1,y_1,z,w)=(0,0,0,w)$. 
The computation is similar. We find a singularity in the chart $2_{zz}$,
while the blown-up equations are regular for the charts $2_{zx}$ and $2_{zy}$.
Here we show the result for the relavant charts $2_{zx}$ and $2_{zz}$.
\\
\\
\noindent
\underline{Chart $2_{zx}$}
\beqa
\Phi_z(x_1, x_1 y_2, x_1 z_2,w)&=&x_1^2\Phi_{zx}(x_1,y_2,z_2,w),\n
\Phi_{zx}(x_1,y_2,z_2,w)&=& x_1(z_2-1)z_2-x_1^2(z_2-1)^3+w-y_2^2.\n
    \mbox{$\cal C$$^{\pm}_{p_2}$ in $2_{zx}$}&:& x_1=0,~~y_2=\pm \sqrt{w}.\n
    \mbox{$\delta$$_{p_2}$ in $2_{zx}$}&:& x_1=0,~~y_2=0.\n
   \mbox{Singularities}&:& \mbox{None}.
\label{2zx}
\eeqa
\\
\noindent
\underline{Chart $2_{zz}$}
\beqa
\Phi_z(x_2 z, y_2 z, z,w)&=&z^2\Phi_{zz}(x_2,y_2,z,w),\n
\Phi_{zz}(x_2,y_2,z,w)&=&
w x_2^2+(x_2-1) z \left(x_2^2 z-2 x_2 z-x_2+z\right)-y_2^2
.\n
  %
   \mbox{$\cal C$$^{\pm}_{p_2}$ in $2_{zz}$}&:& z=0,~~y_2=\pm \sqrt{w} x_2.\n
   \mbox{$\delta$$_{p_2}$ in $2_{zz}$}&:& z=0,~~y_2= 0.\n
   \mbox{Singularities}&:&(x_2,y_2,z)=(0,0,0).
\label{}
\eeqa
\\
\noindent
{\bf 3rd blow up} \\
We finally blow up the codimension-one singularity $(x_2,y_2,z)=(0,0,0)$ 
in the chart $2_{zz}$. It turns out that this completes the resolution process 
completely without leaving any singularities.

The equations of the exceptional curve (with a definite $w$) in the 
relevant charts are:\\
\\
\noindent
\underline{Chart $3_{zzx}$}
\beqa
\Phi_zz(x_2, x_2 y_3, x_2 z_3, w)&=&x_2^2\Phi_{zzx}(x_2,y_3,z_3,w),\n
\Phi_{zzx}(x_2,y_3,z_3,w)&=&
w+(x_2-1) z_3 \left((x_2-1)^2 z_3-1\right)-y_3^2.\n
    \mbox{$\cal C$$_{p_3}$ in $3_{zzx}$}&:&x_2=0,~~y_3^2=w-(z_3-1) z_3.\n 
   \mbox{$\delta$$_{p_3}$ in $3_{zzx}$}&:&x_2=0,~~y_3^2=-(z_3-1) z_3.\n
   \mbox{Singularities}&:&\mbox{None}.
\label{}
\eeqa
\noindent
\underline{Chart $3_{zzz}$}
\beqa
\Phi_zz(x_3 z, y_3 z, z, w)&=&z_2^2\Phi_{zzz}(x_3,y_3,z,w),\n
\Phi_{zzz}(x_3,y_3,z,w)&=&
x_3^2 (w-z (3 z+1))+x_3^3 z^3+3 x_3 z+x_3-y_3^2-1=0.\n
    \mbox{$\cal C$$_{p_3}$ in $3_{zzz}$}&:&z_2=0,~~y_3^2=w x_3^2+x_3-1.\n
    \mbox{$\delta$$_{p_3}$ in $3_{zzz}$}&:&z_2=0,~~y_3^2=x_3-1.\n
   \mbox{Singularities}&:&\mbox{None}.
\label{}
\eeqa

This completes the blowing-up process, and the space is now smooth.
We have seen that conifold singularities
do not appear at any stage of the blow up at the $D_6$ points. 
This is similar to the case of the incomplete resolution 
at the $E_6$ point in the split $I_6$ model.  However, unlike that case, 
the intersection of the exceptional curves does not change at all at the 
$D_6$ points, as we will see in the next section.

\subsection{Intersections of the exceptional curves}
\begin{figure}[htb]
  \begin{center}
         \includegraphics[clip, width=7.6cm]{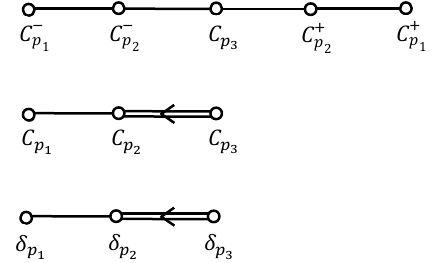}
                \caption{Intersection diagrams of the exceptional curves:
                (Top) $w\neq 0$ before the projection \eqref{eq:A5toC3}; 
                (Middle) $w\neq 0$ after the projection \eqref{eq:A5toC3}; 
                (Bottom) $w=0$.}
    \label{fig:Sp3}
  \end{center}
\end{figure}

At fixed $w\neq 0$, we have five exceptional curves $\cal C$$^{\pm}_{p_1}$, $\cal C$$^{\pm}_{p_2}$
and $\cal C$$_{p_3}$. 
From the above explicit forms, one finds that their intersection matrix is given by 
the $A_5$ Dynkin diagram (the top diagram of Figure \ref{fig:Sp3}).
Although $\cal C$$^{\pm}_{p_1}$ and $\cal C$$^{\pm}_{p_2}$ are
respectively factorized into two lines on this fixed $w\neq 0$ plane, they do not factor 
in the polynomial ring  of $w$.
The two lines at some fixed $w\neq 0$ are interchanged with each other at 
$w=0$, meaning that this is a non-split type of the singularity. Thus the two lines for 
$\cal C$$^{\pm}_{p_1}$ or $\cal C$$^{\pm}_{p_2}$ at fixed $w\neq 0$ comprising the Kodaira 
fibers of type $I_6$ are identified.
Hence we define
\beq
  {\cal C}_{p_i} \equiv \frac{1}{2}({\cal C}^{+}_{p_i} + {\cal C}^{-}_{p_i})  ~~~~ (i=1,2),
\label{eq:A5toC3}
\eeq
which are the projections onto the components invariant  under the diagram automorphism 
of the $A_5$ Dynkin diagram.
Then one can show that the three exceptional curves $\cal C$$_{p_1}$, $\cal C$$_{p_2}$ and $\cal C$$_{p_3}$
form a non-simply-laced Dynkin diagram of $C_3$ (the middle diagram of Figure \ref{fig:Sp3}).

At $w=0$, we again encounter another difference between the
present non-split case and the previous examples of singularities 
associated with the magic square. In the incomplete resolutions 
for the previous examples  
$(G,H)=(E_6,SU(6))$, $(E_7,SO(12))$ and $(E_8,E_7)$, 
while the number of the exceptional fibers at $w=0$ is 
the same as that at $w\neq 0$, 
some of the exceptional fibers at $w=0$ turn out to be
linear combinations of those at $w\neq 0$. Therefore, the intersection 
diagram of the exceptional fibers at $w=0$ becomes different from 
that at $w\neq 0$ 
as we summarized in section 2.2.
Here, we see something different. 
As in the previous works, by lifting up the exceptional curves from the defining chart into 
subsequent charts and seeing their relations, 
one finds that
\beq
  {\cal C}^{\pm}_{p_1} \rightarrow \delta_{p_1},~~ {\cal C}^{\pm}_{p_2} \rightarrow \delta_{p_2},~~ 
  {\cal C}_{p_3} \rightarrow \delta_{p_3}.
\eeq
Substituting them into \eqref{eq:A5toC3}, we obtain 
\beqa
{\cal C}_{p_1} \rightarrow \delta_{p_1}, ~~{\cal C}_{p_2}\rightarrow \delta_{p_2}, ~~
{\cal C}_{p_3}\rightarrow \delta_{p_3}.
\eeqa
Thus, the intersection matrix remains identical even at the codimension-two point
(see the bottom diagram of Figure \ref{fig:Sp3}).
This is a sharp contrast to the previous examples, where the 
intersection matrices at $w=0$ did not coincide with any of (the minus of) 
the Lie algebra Cartan matrices. 


\section{Split/non-split transition as a conifold transition}
In the previous section, we have seen that there is no sign of 
local matter fields near the $D_6$ points.
In this section, 
we will use the recent result of \cite{KuMT} 
to illustrate how the matter fields are considered to arise 
near the $D_6$ points in the non-split $I_6$ model. 
In a nutshell, what has been found in \cite{KuMT} is 
that a transition from the split to the non-split model in F-theory 
is in most cases a transition from the deformed side 
to the resolved side in the conifold transition associated with 
the conifold singularities which arise at $D_{2k}$ points 
(or $E_7$ points for the non-split $IV^*$, which are irrelevant here).
In the present case, they are $D_6$ points, so they are 
precisely what we have been considering in the previous sections.

If we consider the resolution of the split $I_6$ model 
instead of the non-split one, we find various conifold singularities
(Figure 2).
 \begin{figure}[htb]
  \begin{center}
         \includegraphics[clip, width=15cm]{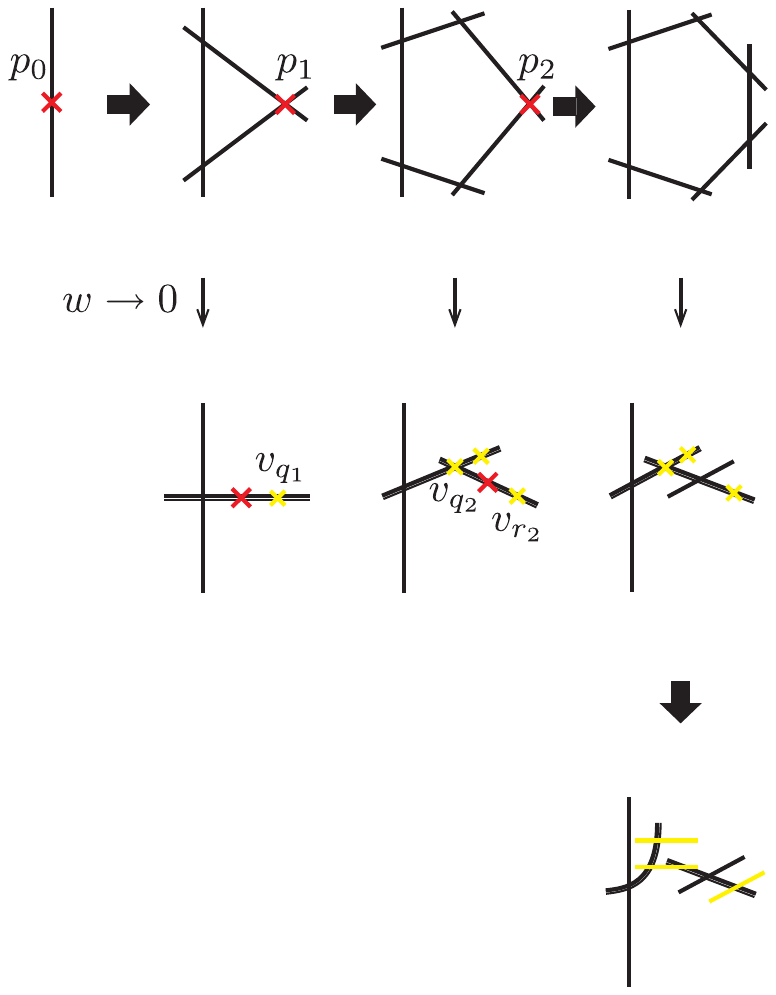}
                \caption{}
    \label{fig:I6blowup}
  \end{center}
\end{figure}
Indeed, by replacing $w$ with $w^2$ in (\ref{1x}), we find 
\beqa
\Phi_x(x,y_1,z_1,w^2)&=&
w^2-x^4 z_1^6+3 x^3 z_1^4+x^2 (z_1-3) z_1^2-x z_1+x-y_1^2\n
&=&
-y_1^2+w^2-x (z_1+O(x)),
\label{1xsplit}
\eeqa
which shows that 
\beqa
v_{q_1}:~(x,y_1,z_1,w)=(0,0,0,0) 
\eeqa
is a conifold singularity. Also, in (\ref{2zx}), $\Phi_{zx}(x_1,y_2,z_2,w^2)$ 
becomes 
\beqa
\Phi_{zx}(x_1,y_2,z_2,w^2)&=& x_1(z_2-1)z_2-x_1^2(z_2-1)^3+w^2-y_2^2\n
&=&-y_2^2+w^2 +x_1\left((z_2-1)z_2+O(x_1)\right),
\label{2zxsplit}
\eeqa
showing that 
\beqa
v_{q_2}:~(x_1,y_2,z_2,w)=(0,0,0,0)~~~\mbox{and} ~~~
v_{r_2}:~(x_1,y_2,z_2,w)=(0,0,1,0)
\eeqa
are conifold singularities.
In this case, it can be shown that the exceptional curves arising from 
their small resolutions precisely yield (together with 
the ones coming from the codimension-one 
singularities) the $D_6$ 
Dynkin diagram as their intersection diagram (Figure 2).

In both the split and non-split cases, 
we can say that the $D_6$ point are 
where  $h_{2n-2r+4}$ vanishes, 
and the split case is when $h_{2n-2r+4}$ 
is in the special form $h_{n-r+2}^2$.
In other words, 
in the split model, a $D_6$ point is a double root of the equation 
$h_{2n-2r+4}=0$, whereas in the non-split model, it is a single 
root. So suppose that $h_{2n-2r+4}=w^2$ near $w=0$ 
in the split case. Then, by a deformation of the complex structure
$w^2\rightarrow w^2-\epsilon^2=(w+\epsilon)(w-\epsilon)$ for some 
small deformation parameter $\epsilon$,  
the double zero $w=0$ becomes a pair of single roots 
$w=\pm \epsilon$, and the split model becomes 
a non-split model accordingly. On the other hand, 
as we can see in eqs. (\ref{1xsplit}) and (\ref{2zxsplit}),
changing $w^2$ to $w^2-\epsilon^2$ 
is exactly turning a conifold into a deformed conifold.
Therefore, 
we see that, at the stage where we have finished blowing up  
all the codimension-one singularities and only conifold singularities remain, 
what we get by a small resolution is a split model, 
and what we get by a deformation is a non-split model. 
In other words, the split/non-split transition is a conifold transition
\cite{KuMT}.

Once this fact is revealed, it is not surprising that 
the conifold singularity does not appear in the non-split model.
Since the non-split model corresponds to a deformed conifold, 
the two-cycles in the split model that are responsible 
for the matter generation are replaced by three-cycles 
in the non-split model.

How do these three-cycles give rise to massless matter fields? 
In \cite{KuMT}, we have discussed several possibilities. 
One of them is the wrapped M5-branes around $S^2\times S^3$.
Since the massless matter in the split model is accounted for 
by the wrapped M2-branes around the vanishing two-cycles, 
this would be a natural guess. The total volume of $S^2\times S^3$ 
will vanish at the apex of the deformed conifold as the volume 
of $S_2$ vanishes there. Also it must contain at least one dimension 
of the elliptic fiber, for which a small volume limit is taken in 
the F-theory limit.  We cannot say anything conclusive in this paper, 
so we leave the clarification of the precise mechanism as an issue 
for the future.

\section{Conclusions}

Motivated by the coincidence between the three examples of 
half-hypermultiplets and the entries of the magic square, 
we have studied a six-dimensional ${\cal N}=1$ 
F-theory compactification on an elliptic 
fibration over a Hirzebruch surface with a codimension-one 
singularity of the {\em non-split} $I_6$ type found in \cite{BIKMSV}. 
This model supports an $Sp(3)$ gauge symmetry. 
The heterotic index and the generalized Green-Schwarz analysis  
both show that such a compactification gives massless 
half-hypermultiplets in the ${\bf 14}'$ representation (as well as 
the ${\bf 6}$ reprentation) of $Sp(3)$, 
which is $F_4/(Sp(3)\times SU(2))$ 
($Sp(4)/(Sp(3)\times SU(2))$). 
We have shown that they are generated at the $E_6$ points,
where half-hypermultiplets ${\bf 20}$
of $SU(6)$ would have appeared in the split model. 
In the non-split model, $SU(6)$ is broken to $Sp(3)$, and 
${\bf 20}$ is decomposed into ${\bf 14}'\oplus{\bf 6}$ of $Sp(3)$ 
accordingly, yielding the desired multiplets.

We have also considered the problem on the non-local matter 
generation near the $D_6$ point. We have pointed out 
two puzzles: The first one is how the degrees of massless matter 
fields in the split model can be plausibly assigned at the zero loci 
of the relevant section  $h_{2n+4-2r}$, the number of which is 
doubled in the transition from the split to non-split models.  
Second, by performing a singularity resolution, 
we found no indication of the existence of localized massless 
matter fields. 
We have explained why this is so by using the result of \cite{KuMT} that
the split/non-split transition can be regarded as a 
conifold transition.

\section*{Acknowledgement}
We thank Y.~Kimura and H.~Otsuka for useful discussions.



\begin{thebibliography}{99}

\bibitem{Vafa}
C.~Vafa,
  Nucl.\ Phys.\ B {\bf 469}, 403 (1996)
  [hep-th/9602022].


\bibitem{MV1}
D.~R.~Morrison and C.~Vafa,
  Nucl.\ Phys.\ B {\bf 473}, 74 (1996)
  [hep-th/9602114].

\bibitem{MV2}
D.~R.~Morrison and C.~Vafa,
  Nucl.\ Phys.\ B {\bf 476}, 437 (1996)
  [hep-th/9603161].


\bibitem{BIKMSV} M.~Bershadsky, K.~Intriligator, S.~Kachru, D.R.~Morrison, V.~Sadov and C.~Vafa,
Nucl.Phys. B481 (1996) 215-252 [hep-th/9605200].

\bibitem{KatzVafa}
S.~H.~Katz and C.~Vafa,
  Nucl.\ Phys.\ B {\bf 497}, 146 (1997)
  [hep-th/9606086].

\bibitem{Tani}
T.~Tani,
  Nucl.\ Phys.\ B {\bf 602}, 434 (2001).


\bibitem{Curio} 
  G.~Curio,
  Phys.\ Lett.\ B {\bf 435}, 39 (1998)
  [hep-th/9803224].

\bibitem{DiaconescuIonesei} 
  D.~E.~Diaconescu and G.~Ionesei,
  JHEP {\bf 9812}, 001 (1998)
  [hep-th/9811129].

\bibitem{MTanomaly}
S.~Mizoguchi and T.~Tani,
PTEP \textbf{2016} (2016) no.7, 073B05
[arXiv:1508.07423 [hep-th]].

\bibitem{MTLooijenga}
S.~Mizoguchi and T.~Tani,
JHEP \textbf{11} (2016), 053
[arXiv:1607.07280 [hep-th]].

\bibitem{OguisoShioda}
K. Oguiso and T. Shioda, 
Comment. Math. Univ. St. Pauli. 40 (1991) 83.




\bibitem{Wolf}
J. A. Wolf, 
J. of Math. Mech., 14 (1965), 1033.

\bibitem{Alekseevskii}
D.V. Alekseevskii, 
Funct. Anal. Appl. 2 (1968), 97; 
Funct. Anal. Appl. 2 (1968), 106; 
Math. USSR  Izv. 9 (1975), 297.

\bibitem{Dasguptaetal}
K.~Dasgupta, V.~Hussin and A.~Wissanji,
Nucl. Phys. B \textbf{793} (2008), 34-82
[arXiv:0708.1023 [hep-th]].


\bibitem{MT}
  D.~R.~Morrison and W.~Taylor,
  JHEP {\bf 1201}, 022 (2012)
  [arXiv:1106.3563 [hep-th]].

\bibitem{KMT}
N.~Kan, S.~Mizoguchi and T.~Tani,
[arXiv:2003.05563 [hep-th]]. To appear in JHEP.


\bibitem{FFamilyUnification}
 S.~Mizoguchi,
  JHEP {\bf 1407}, 018 (2014)
  [arXiv:1403.7066 [hep-th]].


\bibitem{KugoYanagida} 
  T.~Kugo and T.~Yanagida,
  Phys.\ Lett.\  {\bf 134B}, 313 (1984).
  

\bibitem{GHLST}
A.~Grassi, J.~Halverson, C.~Long, J.~L.~Shaneson and J.~Tian,
JHEP \textbf{09} (2018), 129
[arXiv:1805.06949 [hep-th]].

\bibitem{AGW}
P.~Arras, A.~Grassi and T.~Weigand,
J. Geom. Phys. \textbf{123} (2018), 71-97

\bibitem{EJK}
M.~Esole, P.~Jefferson and M.~J.~Kang,
[arXiv:1704.08251 [hep-th]].

\bibitem{EK}
M.~Esole and M.~J.~Kang,
JHEP \textbf{02} (2019), 091
[arXiv:1805.03214 [hep-th]].

\bibitem{EJ}
M.~Esole and P.~Jefferson,
[arXiv:1910.09536 [hep-th]].


\bibitem{KuMT}
R.~Kuramochi, S.~Mizoguchi and T.~Tani,
[arXiv:2108.10136 [hep-th]].


\bibitem{magical1}
  M.~Gunaydin, G.~Sierra and P.~K.~Townsend,
  Phys.\ Lett.\ B {\bf 133} (1983) 72.

\bibitem{magical2} 
M.~Gunaydin, G.~Sierra and P.~K.~Townsend,
  Nucl.\ Phys.\ B {\bf 242} (1984) 244.

\bibitem{Kanmagical}
N.~Kan and S.~Mizoguchi,
Phys. Lett. B \textbf{762} (2016), 177-183
[arXiv:1605.01904 [hep-th]].

\bibitem{Fukuchimagical}
S.~Fukuchi and S.~Mizoguchi,
Phys. Lett. B \textbf{781} (2018), 77-82
[arXiv:1802.06555 [hep-th]].


\bibitem{GSW}
M.~B.~Green, J.~H.~Schwarz and P.~C.~West,
Nucl. Phys. B \textbf{254} (1985), 327-348

\bibitem{Yamatsu}
N.~Yamatsu,
[arXiv:1511.08771 [hep-ph]].


\bibitem{Sadov}
V.~Sadov,
Phys. Lett. B \textbf{388} (1996), 45-50
[arXiv:hep-th/9606008 [hep-th]].


\bibitem{MizoguchiTanianomaly}
S.~Mizoguchi and T.~Tani,
PTEP \textbf{2016} (2016) no.7, 073B05
[arXiv:1508.07423 [hep-th]].


\bibitem{BoxGraphs}
  H.~Hayashi, C.~Lawrie, D.~R.~Morrison and S.~Schafer-Nameki,
  JHEP {\bf 1405}, 048 (2014)
  [arXiv:1402.2653 [hep-th]].

\bibitem{MTnonCartan}
S.~Mizoguchi and T.~Tani,
JHEP \textbf{03}, 121 (2019)
[arXiv:1808.08001 [hep-th]].



\end{thebibliography}
\end{document}